\documentclass[%
	amsmath,
	amssymb,
	aps,
	showpacs,title
	showkeys,
	twocolumn,
	altaffilletter,
	nolongbibliography,
	numerical,
	flushbottom,
	secnumarabic,
	prl,
	superscriptaddress,
	floatfix
]{revtex4-1}

\usepackage{lipsum}

\usepackage[dvipsnames]{xcolor}

\usepackage{bm}

\usepackage{graphicx}

\usepackage{subfigure}

\usepackage{upgreek}

\usepackage{natbib}

\usepackage{braket}

\usepackage{physics}

\usepackage{siunitx}

\usepackage[percent]{overpic}

\usepackage{color}

\begin{document}

\author{V.~Vitale}
\affiliation{Dipartimento di Fisica "Ettore Pancini", Universit\`a di Napoli ``Federico II'', Monte S.~Angelo, I-80126 Napoli, Italy}

\author{G.~De Filippis}
\affiliation{Dipartimento di Fisica "Ettore Pancini", Universit\`a di Napoli ``Federico II'', Monte S.~Angelo, I-80126 Napoli, Italy}
\affiliation{CNR-SPIN, Monte S.~Angelo via Cinthia, I-80126 Napoli, Italy}

\author{A.~De Candia}
\affiliation{Dipartimento di Fisica "Ettore Pancini", Universit\`a di Napoli ``Federico II'', Monte S.~Angelo, I-80126 Napoli, Italy}

\author{A.~Tagliacozzo}
\affiliation{Dipartimento di Fisica "Ettore Pancini", Universit\`a di Napoli ``Federico II'', Monte S.~Angelo, I-80126 Napoli, Italy}
\affiliation{CNR-SPIN, Monte S.~Angelo via Cinthia, I-80126 Napoli, Italy}

\author{V.~Cataudella}
\affiliation{Dipartimento di Fisica "Ettore Pancini", Universit\`a di Napoli ``Federico II'', Monte S.~Angelo, I-80126 Napoli, Italy}
\affiliation{CNR-SPIN, Monte S.~Angelo via Cinthia, I-80126 Napoli, Italy}

\author{P.~Lucignano}
\email[]{procolo.lucignano@spin.cnr.it}
\affiliation{CNR-SPIN, Monte S.~Angelo via Cinthia, I-80126 Napoli, Italy}
	\affiliation{Dipartimento di Fisica "Ettore Pancini", Universit\`a di Napoli ``Federico II'', Monte S.~Angelo, I-80126 Napoli, Italy}

\title{Assessing the quantumness of the annealing dynamics via Leggett Garg’s inequalities: a weak measurement approach
}

\date{\today}

\maketitle

{\bf Adiabatic quantum computation (AQC) ~\cite{albash:review-aqc} is a promising counterpart of universal quantum computation \cite{nielsen:quantum-computation},  based on the key concept of quantum annealing (QA)~\cite{kadowaki:qa}. 
 QA is claimed to be at the basis of commercial quantum computers \cite{D-wave:2010} and benefits from the fact that  the detrimental role of decoherence and dephasing  seems to have poor impact on the annealing towards the ground state.
 While many papers \cite{Johnson10,Harris10,King15} show interesting optimization results with a sizable number of qubits, a clear evidence of a full quantum coherent behavior during the whole annealing procedure is still lacking.
In this paper we show that  quantum non-demolition (weak) measurements \cite{Aharonov:PRL1988} of  Leggett Garg inequalities  can be used to  efficiently assess the quantumness of  the  QA  procedure. Numerical simulations based on a weak coupling Lindblad approach \cite{zanardi:master-equations} are compared with classical Langevin simulations to support our statements.}

Although any quantum algorithm can be run on adiabatic quantum computers~\cite{Das:RMP2008},  the interest of the scientific community is focused on decision and optimization problems that are very difficult to handle on classical computers, because their computational time, {most of the times},  grows exponentially with the number of bits. 
Optimization problems can be mapped onto complex many body hamiltonians~\cite{lucas:np-problems}, hence AQC is also of outmost interest from the fundamental  point of view, as it may provide insights into longstanding problems in modern condensed matter physics  such as physics of the strongly correlated cuprate materials \cite{lee2006doping} and of the spin glasses   \cite{barahona1982computational}. 

AQC {is founded on} QA, a slow quantum dynamics that proceeds from an initial Hamiltonian with a trivial ground state (easy to prepare), to a final Hamiltonian whose ground state encodes the solution of the computational problem. The adiabatic theorem guarantees that the system will track the instantaneous ground state if the Hamiltonian varies sufficiently slowly~\cite{farhi:quantum-computation}. 
QA can perform better than thermal annealing~\cite{santoro:spin-glass,martonak:salesman}, 
however, to the date, there are only a few problems where this quantum speed-up has been clearly demonstrated.
\\
Furthermore, in the last few years, there has been a renewed interest in QA~\cite{harris:d-wave, ronnow:quantum-speed-up, boixo:experimental-signature,boixo:hundred-qubits,shin:d-wave},
because physical {superconducting} implementations of "quantum" annealers~\cite{Johnson10,Harris10,King15} up to thousand of spins have been used to obtain the GS of interacting many body Hamiltonians. {Moreover also optical techniques have been successfully used \cite{Lechner:2015,Lechner:2017}, though with a smaller number of qubits.}
In real devices, the presence of the environment introduces decoherence and thermalization time scales that have to be longer than the annealing time ~\cite{zanardi:master-equations}. This condition does not strictly guarantee that the dynamics can be assumed to be quantum for the whole evolution. Moreover, as the presence of a classicizing environment is unavoidable, the questions we want to address in the present letter are the following: 
\begin{itemize}
\item[] To what extent, the QA  can be considered a quantum coherent dynamics, in the presence of a dissipative environment?
\item[] Once we choose an annealing time $t_f$ at which a certain target state is (almost) reached, is it possible to find a "quantumness" estimator of the evolution towards such state? 
\end{itemize}
Giving a clear answer to the previous questions, is not only a matter of semantics. 
It is well known that  quantum speed up \cite{nielsen:quantum-computation} can only be accessed if quantum mechanical coherence is preserved during the whole dynamics.

In order to answer to these questions, we propose to evaluate the Leggett-Garg's inequalities (LGI) during the QA. 
LGI were developed in 1985, to study quantum coherence properties of macroscopic quantum systems~\cite{Leggett:PRL_1985}, {that have been recently studied to assess the quantumness of a damped two level system \cite{Friedenberger:2017}.}
They are Bell's-like inequalities in time, and predict anomalous values for some correlation functions that are only possible if the system behaves according to quantum mechanics. They also provide sharp bounds for classical correlation functions~\cite{Emary:ROP_2014}.\\

In this paper we focus on a simple model i.e. we study the QA of a single qubit. 

{Of course, the dynamics of a single qubit does not contain all the relevant ingredients of an interacting multi-qubit system.  Phenomena like  many-body tunneling, entanglement, and many body localization are, of course, not included.  Although the generalization of the proposed  approach to multi-qubits is straightforward (see Supplementary Information) we believe that the simple single-qubit case is the fundamental first step for understanding how quantum mechanical coherent behaviors are spoiled by the system-bath interaction, even if  it neglects the aforementioned many-qubit phenomena.  However, even if a full analisys is beyond the aim of the present work, in the Supplementary Information we  show that the proposed procedure is still meaningful for two interacting qubits providing that the coupling is not very strong. 
Hence, in the following, for sake of clarity, we give a detailed description of the physics of the single qubit.\\
The  QA is}  described by the following time-dependent Hamiltonian:
\begin{equation}
H(s)=(1-s)\frac{\Gamma_x}{2}\sigma_x +s \frac{\Gamma_z}{2}\sigma_z.
\label{ham0}
\end{equation}
where $s=t/t_f,s \in [0,1]$, $t_f$ is so-called annealing time and $\sigma_i \;({i=x,z})$ are the Pauli matrices describing the qubit as a quantum two level system. At the initial time $s=0$ the system is prepared into the GS that is the  $\sigma_x$ eigenstate $|GS(0)\rangle =1/\sqrt{2}(|\uparrow\rangle-|\downarrow\rangle)$, and eventually at the final time $s=1$  it is annealed {towards	}  $|GS(1)\rangle =|\downarrow\rangle$. Of course this single qubit QA does not entail the complexity of a many qubit problem, however it is a simple prototypical model to address decoherence and relaxation phenomena   out of equilibrium.
Following Ref.~\cite{albash:decoherence}, we set our time/energy scale choosing $\Gamma_x=\Gamma_z=1\text{GHz}$ (that is the typical working frequency of the experimentally relevant annealers based on superconducting flux qubits \cite{D-wave:2010})  and express all the energies in units of $\Gamma_x$  (times in units of $1/\Gamma_x$, $\hbar=1$).
The qubit environment is described by a bath of bosonic harmonic oscillators in thermal equilibrium at the inverse temperature $\beta=1/k_BT$ ($k_B$ is the Boltzmann constant). The qubit bath coupling is described by an ohmic spectral density whose effective interaction strength is the dimensionless parameter $\alpha$. At $\alpha=1$ (for $s=0$) the system undergoes the Leggett transition \cite{leggett1987dynamics}, however in this paper we will focus on the weak coupling limit ($\alpha \ll 1$) far away from this critical point. For a detailed description of the system bath interaction please refer to the Supplementary Information.

%
%
The reduced density matrix $\rho_Q$ describing the qubit only is obtained by tracing over the environment degrees of freedom. We adopt a  Lindblad approach (for details see Ref.s~\cite{breuer:open-quantum,zanardi:master-equations})  that gives the following master equation for the density matrix:
\begin{equation}
\frac{d\rho_Q(t)}{dt}=-i[H(t)+H_{LS}(t),\rho_Q(t)]+\mathcal{D}_t[\rho_Q(t)],
\end{equation}
where $H_{LS}(t)$ is the Lamb shift Hamiltonian and $\mathcal{D}_t$ is the dissipator, responsible for the non unitary dynamics.
They are both described in terms of local (in time) Lindblad operators $L(t)$. A detailed description of $H_{LS}(t)$, $\mathcal{D}_t$ and the Lindblad operators is given in the Supplementary Information.\\
\begin{figure}
	\centering
	\begin{overpic}[width=\linewidth]{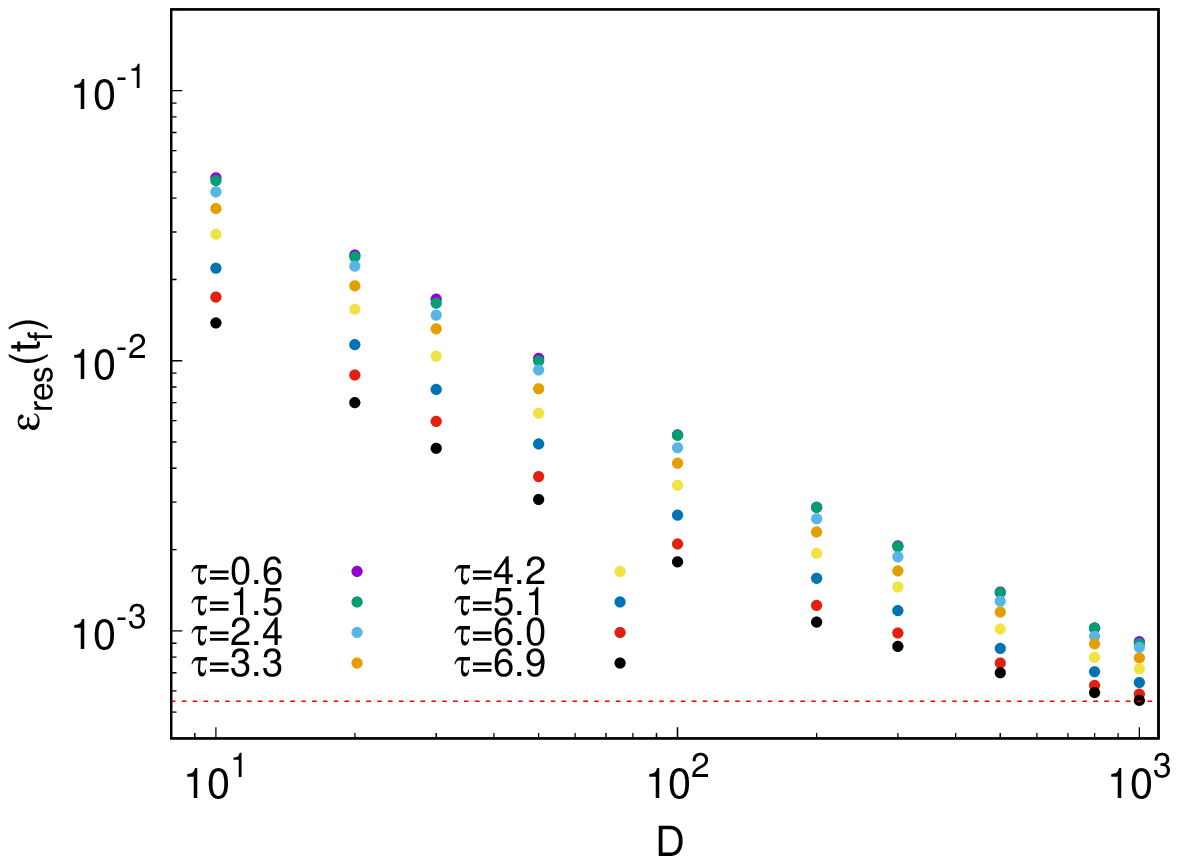}
		\put(45,44){\includegraphics[width=0.5\linewidth]{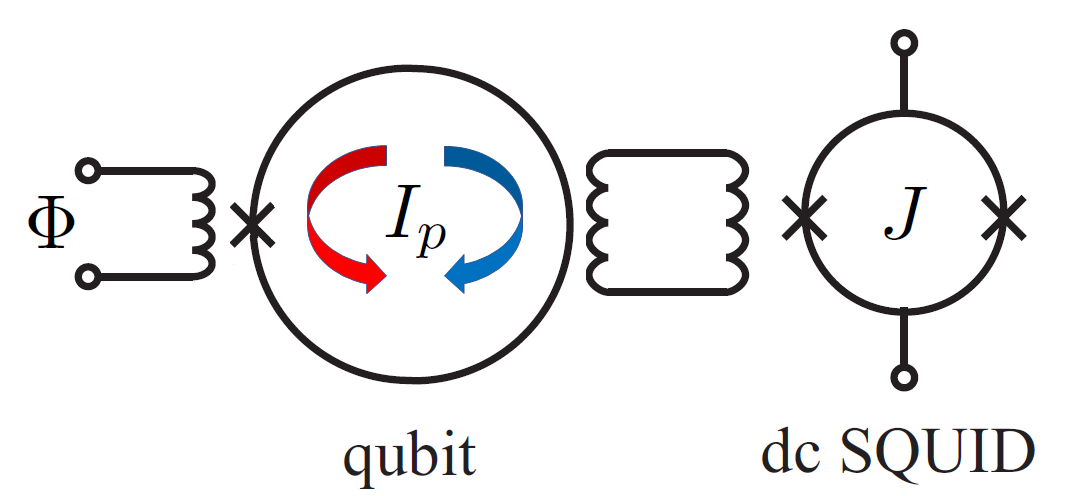}} 
	\end{overpic}
	\caption[]{Log-log plot of the residual energy of the system as a function of the variance $D$ and of the times at which the measurements are performed: $t_1=\tau$ and $t_2=2\tau$. We set $t_f=14$. We observe that the residual energy decreases eventually going to $5.49 \cdot 10^{-4}$ which is the residual energy of the system in the absence of measurements. The inset in the right top corner shows a diagram of the system-ancilla ensemble.} \label{fig:residualenergyDintro}
\end{figure}
In this paper we focus on two estimators. 

{\it The residual energy} $\epsilon_{\text{res}}$  tells us  whether our adiabatic dynamics is successful (or not) in reaching the target state and is defined as the difference between the energy of the system at the final time $t_f$ and the exact ground state $E_0(t_f)$ of the target Hamiltonian $H(t_f)$. 
\begin{equation}
\epsilon_{\text{res}}=Tr [\rho_Q H(t_f)] - E_0(t_f),
\end{equation}
 Of course, due to the adiabatic theorem~\cite{morita2008mathematical}  $\epsilon_{\text{res}}$ tends to zero if $t_f \rightarrow \infty$, {if the evolution is  unitary}.\\

{\it The Leggett-Garg's correlation functions} tell us if the system behaves quantum coeherently during its dynamics \cite{Emary:ROP_2014}.
We focus on the third-order Leggett-Garg's function:
\begin{eqnarray}
K_3^a&=& C_{12}+C_{23}-C_{13}\\
C_{i,j}&=& \langle \sigma_z(t_i)\sigma_z (t_j)\rangle \nonumber
\end{eqnarray}
and other nonequivalent third order functions $K^{b}_3=-C_{12}-C_{23}-C_{13}$, and $K^{c}_{3}=-C_{12}+C_{13}+C_{13}$  obtained by nontrivial cycling of the $1,2,3$ indexes (all the other permutations are trivially reduced to one of these three). 
If the system behaves classically then:
\begin{equation}
-3\leq K_3^i \leq 1\:,\; i\in\{a,b,c\};.
\end{equation}
Hence, in the following, we seek for  violation of Leggett Garg's inequalities during the annealing to make sure that our system behaves quantum mechanically up to the annealing time $t_f$.\\%
\begin{figure}
	\centering
	\includegraphics[width=\linewidth]{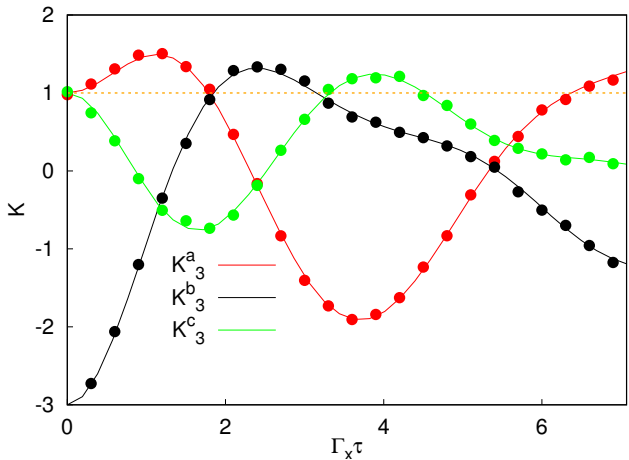}\caption[]{ Plot of the Leggett-Garg's function, in the absence of coupling to the environment. The lines are obtained performing projective measurements. The dots are calculated with weak measurements considering $N=10^6$, $D=20$. $K^a_3$ in red, $K^b_3$ in black and $K^c_{3}$ in green. The orange line marks the bound of the LGI. Here $\tau= t_2-t_1=t_3-t_2$, $\tau\in [0, t_f/2]$.} 
	\label{fig:LGI}
\end{figure}
Studying the LGI requires the evaluation of two times correlation functions, which implies measuring the system twice during the annealing dynamics.  Conventional projective measurements~\cite{nielsen:quantum-computation}  are  detrimental for the adiabatic quantum computation, indeed after the measurement the system may  populate instantaneously many excited states (see Supplementary Information) reducing the fidelity very close to zero. By contrast, we adopt the paradigm of weak measurements \cite{Aharonov:PRL1988},  to gain relevant information from  the LGI with negligible effect on the annealing results.\\
In particular we propose adopting the measurement approach designed in Ref.~\cite{picot2010quantum} to weakly measure the  LGI. 
The qubit described by a pseudospin degree of freedom $\sigma_z$ is coupled to an ancilla. Measuring the ancilla we can extract information on the qubit state. The stronger the coupling between the two systems, the larger the influence on the quantum annealing dynamics (for a detailed discussion on the measurement scheme see  Supplementary Information).
In order to specify an experimentally accessible weak measurement setup, following Ref.~\cite{picot2010quantum},  we describe a  superconducting flux qubit (based, for instance on a RF SQUID) inductively coupled to a hysteretic DC SQUID acting as an ancilla device as shown in the inset of  Fig. ~\ref{fig:residualenergyDintro}.  The ancilla is biased close to its critical current via a short current pulse. The interaction between the two loops is given by  $H_I = M I_p \sigma_z J$ where $I_p$ is the current circulating in the qubit, J the current in the ancilla and M the mutual inductance coefficient.  At this point the ancilla has a certain probability of switching to the dissipative state, depending on the value of $\sigma_z$, hence on the qubit state.  
We propose to measure the output voltage of the ancilla SQUID, which hence is intimately connected to the qubit state.
If the ancilla relaxation time is shorter than the, so called, discrimination time (for which the signal to noise ratio is close to unity \cite{lupacscu2004nondestructive}), one cannot determine with certainty which is the output voltage, then $\sigma_z$. Therefore one may consider the value of the voltage as distributed according to the following probability distribution $P(V)$ as in \cite{WilliamsJordan:PRL2008}:
\begin{equation}
P(V)=\rho_{Q\downarrow\downarrow} P_-(V) + \rho_{Q\uparrow\uparrow}P_+(V).
\end{equation}
Here $V$ is binormally distributed, peaked around the two values it could assume in the case of projective measurements, say $V_0$ and $\bar{V}$ ($P_\pm$ are the gaussians distributions centred around the two voltage states, $V_0$ and $\bar {V}$). Then the qubit state $\sigma_z$ can be derived from the output $V$ and  will be distributed according to P($\sigma_z$) as well, peaked around (-1,1), with a variance $D$ proportional to the ratio between the the ancilla relaxation time and the discrimination time. The quantities $\rho_{Q\downarrow\downarrow}$ and $\rho_{Q\uparrow\uparrow}$ are the diagonal elements of the qubit  density matrix $\rho_Q$ in the computational basis($\ket{\uparrow},\ket{\downarrow}$).\\
Within this approach, it is also possible to calculate how measuring the ancilla influences the qubit. Indeed, the update rule of the density matrix $\rho_Q$  becomes very simple (see Supplementary Information)  and it is evident that the shorter is the pulse, the less the system is perturbed.
The density matrix $\rho'_Q$ after the measurement is related to the one before the measurement $\rho_Q$ by the update rule \cite{jordan2006qubit,JordanKorotkovButtiker2006}:
\begin{equation}\label{matrixupdateruleintro}
\rho'_Q=\frac{1}{\rho_{Q\downarrow\downarrow}e^{\gamma}+\rho_{Q\uparrow\uparrow}e^{-\gamma}} \begin{pmatrix}
\rho_{Q\downarrow\downarrow} e^{\gamma} & \rho_{Q\downarrow\uparrow} \\
\rho_{Q\downarrow\uparrow}^{*} & \rho_{Q\uparrow\uparrow}e^{-\gamma}\\
\end{pmatrix},
\end{equation}
where $\gamma=\sigma_z(t)/D$.\\ In order to extract meaningful information on the qubit, one must repeat the same evolution many times and evaluate $\sigma_z(t)$ as the average of the different results.\\
This approach allows us to simultaneously measure spin-spin correlation functions, with negligible effect on the annealing dynamics, hence on the residual energy as it will be clearer in the following (detail on our weak measurement scheme in Supplementary Information). \\
\textit { Results and Discussion} \\
In our model hamiltonian, we choose  $t_f=14$ as annealing time. In the unitary limit,  this guarantees that the adiabatic condition is fully satisfied. Adiabaticity is evident by analysing the behaviour of the residual energy as a function of $t/t_f$ at fixed $t_f=14$   and the corresponding ground state population (see Supplementary Information Fig. 1 bottom left panel). 
At the final time the residual energy is very close to zero $\epsilon_{res} = 5.49\times 10^{-4} $ and the fidelity (ground state population at the annealing time) is $\rho_{Q\downarrow\downarrow}(t_f) = 0.999$, when no spin-spin correlation function is "measured" during the annealing dynamics, in  agreemeent with Ref.~\cite{albash:decoherence}.\\
As the Hamiltonian is time dependent, we expect that $C_{i,j}$ will have a generic dependance on the two times $(t_i,t_j)$, $C_{i,j}=C_{i,j}(t_i,t_j)$.
However in the following we study the LGI  as a function of $\tau=t_2-t_1=t_3-t_2$, because maximum violation is expected (in the absence of dissipative baths) when the three times are equally spaced.\\
When evaluating $C_{i,j}$ we perfom two weak measurements per annealing and then calculate their expectation values  averaging over N repeated  dynamics.
The larger the variance $D$, the less the system is affected by the maesurement, and the closer is the residual energy  to its  unperturbed value.\\
To analyse how the measurement procedure affects the dynamics, in Fig. \ref{fig:residualenergyDintro} we show the residual energy for the anneling Hamiltonian of Eq.\eqref{ham0}, with no interaction with the environment, obtained for different choices of the variance $D$, at different values of $\tau$. 
The results are presented repeating the evolution $N=10^6$ times.\\
The residual energy  $\epsilon_{\text{res}}$ at $D=20$ ranges from $7.0\times 10^{-3}$ to $2.5 \times 10^{-2}$ depending on $\tau$. Correspondingly, the fidelity ranges from 0.973 to 0.993. That makes us comfortable about the  convergence of the annealing procedure towards the target state, hence we choose $D=20$ for all the following simulations. \\
\begin{figure*}	
	\includegraphics[width=0.8 \linewidth]{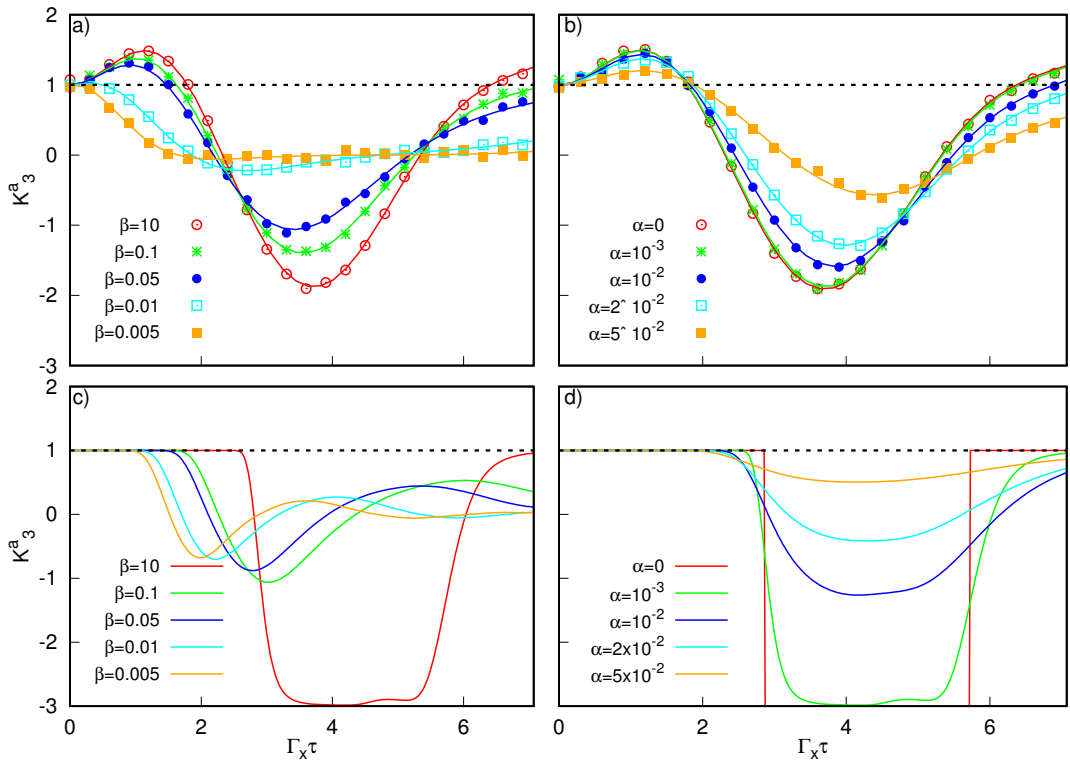}
	\caption[]{Plot of the Leggett-Garg's function $K^a_3$ during the annealing dynamics.
		The black dashed line highlights the upper bound for the LGIs. The LG's functions are plotted as a function of the difference of the times at which the measurements are perfomed: $t_2-t_1=t_3-t_2=\tau$ (in units of $\hbar/ \Gamma_x$ with $\hbar=1$). The time $\tau$ goes from 0 to $t_f$/2 so that it scans the whole evolution ($t_f=14$). Top panels present results of quantum simulations, bottom panels of classical Langevin dynamics described in the Supplementary Information.
		In a) and c) we set $\alpha=10^{-3}$, in b) and d) $\beta=10$.}
	\label{fig:LG_all}
\end{figure*}
In Fig.\ref{fig:LGI} we show the Leggett-Garg's functions during the annealing dynamics in the unitary case, to be compared with the outcomes in the presence of a decoherence bath. The bold lines are obtained in the case of projective measurements, while the dots represent weak measurement results. 
{While in a single "weak" measurement procedure, occasional violations may be driven by the system-detector interaction, averaging over N reapeated measures, assures the convergence to the projective results.}
The agreement between the correlation functions calculated by means of projective and weak measurements demonstrate that our method reproduce the same correlations without affecting in a significant way  the residual energy.  Hence the quantum computation can be successful even during the LGI testing.\\
We clearly see that all the Leggett-Garg's functions go beyond the unitary "classical" limit. $K^b_3$ and $K^c_{3}$ are violated at intermediate times while $K^a_3$ for short and long times.
{Remarkably, at short times, $K^a_3$ and $K^b_3$ show complementary violations, as well as in the case of a single qubit oscillating between its two states, at its own frequency (without annealing) \cite{Emary:ROP_2014}, that is possibly due to the fact that, at short times, the qubit hamiltonian is "mostly" proportional to $\sigma_x$ \cite{Friedenberger:2017}}.
Moreover, at {$\tau=t_f/2=7$} there is a significant violation of $K^a_3$.
This value corresponds to a final measurement time $t_3=t_f=14$. Hence a violation of $K^a_3$ at {$\tau=t_f/2$}, is particularly relevant, because it is a direct proof of quantum coherence until the end of the annealing dynamics. Hence in the following we will focus on the long time {($\tau\sim t_f/2$)} behavior of $K^a_3$.\\
Now we move to discuss the same results in the presence of the dissipative environment. Hence, we turn on the system bath interaction and again calculate the LGIs fixing $D=20$, $t_f=14$. 
The value of $K^a_3$ drops below the classical bound for the LGIs at the final time  both increasing the temperature (see Fig.~\ref{fig:LG_all} a) and increasing the coupling $\alpha$ (see Fig.~\ref{fig:LG_all} b).
That corresponds to a lost quantum coherence at the annealing time $t_f$.\\
It is evident that the temperature plays a key role in the detrimental effect of the thermal bath. For low temperatures the quantum behavior persists during the whole evolution even in the presence a finite of coupling with the environment. By contrast, increasing the temperature, the time during which the system shows quantum features decreases, eventually going to zero for very high temperatures.\\ Since we have used a master equation  in the Lindblad form, we must consider that this approach guarantees reliable results only in the weak coupling limit. Therefore the results shown for very high temperatures, and strong coupling, might be beyond our approximation and have to be considered with care. 
Moreover, at high temperatures, also the description of a SQUID flux-qubit as a two level system ceases to be correct.  
Hence we focus on results at low temperatures ($\beta=10$) and weak-intermediate couplings. In this case (Fig.~\ref{fig:LG_all}b)) the $K^a_3$ function violate the unitary limit at long times from $\alpha=0$ up to  $\alpha=10^{-3}$.  However  increasing the coupling to $\alpha = 10^{-2}$ the value of $K^a_3$ is very close to its "classical" upper bound. Hence, in this case, we decided to simulate the system also using a classical Langevin dynamics to assess the LG functions  in the absence of quantum correlations, as a comparison case (see Supplemetary Information). 
Results are presented in the bottom panels of Fig.s~\ref{fig:LG_all}c), \ref{fig:LG_all}d). 
First of all we notice that, as expected, the LG functions never violates their calssical upper bounds (dashed line).
Interestingly enough, the high temperature results (see Fig.s~\ref{fig:LG_all}a) and c)) show  agreement between classical and quantum dynamics. {Of course we cannot make strong claims comparing Langevin and Lindblad dynamics, however this result points to poor quantum correlations in the Lindblad dynamics at these couplings}. In the case of low temperatures (see Fig.s~\ref{fig:LG_all}b) and d)), at weak coupling we show a completely different behavior of the $K^a_3$ that can be hence used as quantum estimator.  At intermediate couplings, quantum dynamics does not allow for full violation  of LGI, nontheless, the behavior of $K_3^a$ show remarkable and qualitative differences with respect to the classical case. In particular we notice that the ordering of the curves as a function of the coupling at the annealing time (corresponding to $\tau=7$) is reversed between the quantum and the classical case, except for the cases $\alpha =0, 10^{-3}$. Hence, even in these "border line" cases, the evaluation of LGI's could be relevant to asses whether the dynamics has occurred {via} a quantum or a classical path.  A comparative analysis of long time dynamics is shown in the Supplementary Information.  \\

In conclusion, the presence of a dissipative environment can modify the dynamics of a quantum two level system during the QA. {In general, it is detrimental even if, under particular cirumstances,  it may also improve annealing performaces \cite{Passarelli:2018}.}  This does not forbid, in principle, to successfully reach the target state at the annealing time.
The question whether the dynamics has driven the system through a quantum or a classical path, up to now, remained unanswered and only partially addressed in the literature.\\
In this paper we propose a new, powerful and experimentally accessible method to assess the quantumness of a system during its adiabatic evolution, based on the LGI that we evaluate in the framework of weak measurements. \\
It allows us to show that time correlations can be measured  without perturbing the annealing dynamics and that LGIs hold information about the interaction with the environment and can be used as witness of quantum coherence.\\
Do our results allow us to answer the point raised at the beginning of this paper? Namely: are real annealers, claimed to work performing adiabatic quantum computation, really quantum annealers? Are their outcomes macroscopic manifestations of quantum mechanics?\\
Our results show, for a very simple model, that if one measures the LGIs along the adiabatic dynamics, a possible, yet non trivial, outcome could be that  $K^a_3(t_f/2)$  contains all the information we need to  assess if the final result of the computation is quantum or not. 
In the case of long time violation $K^a_3(t_f/2)>1$ the result has to  be considered as quantum, even in the presence of a dissipative bath. In borderline cases $K^a_3(t_f/2)\sim 1$,  a careful analysis of the $K^a_3$ behaviour as a function of time, could unveil the characteristics of the dynamics, namely if it was quantum, classical, or due to a non trivial occurrence of quantum and classical mechanisms.
However this approach is still at its infancy. Extending it to more complicated ensembles like N spin Ising chains would be a fascinating way along which to proceed.\\

\textbf{Data availability}\\
 The data that support the plots within this paper and other findings of this study are available from the corresponding author upon request.\\

\textbf{Competing interests}\\
The authors have no potential financial or non-financial conflicts of interest.\\

\bibliographystyle{IEEEtran}

\textbf{Acknowledgments}\\
The authors acknowledge stimulating discussions with Pino Falci, Rosario Fazio and Giuseppe E. Santoro { as well as Gianluca Passarelli for his invaluable support in optimizing the performances of the Lindblad code.} \\

\textbf{Author contributions}\\
GD, VC, PL, AT set the theoretical framework.
PL and VV wrote the Lindblad code to weakly measure LGI during the QA.
VV performed the quantum simulations.
AD performed the classical-quantum mapping and the classical simulations.
PL wrote the paper, with help from all co-authors.\\

\textbf{Additional Information}\\
Supplementary information accompanies this paper. Reprints and permissions information is available online. Correspondence and requests for materials should be addressed to P.L.

\end{document}